\documentclass{aa}
\usepackage{natbib}
\usepackage{graphicx}
\usepackage{txfonts}
\usepackage{color}
\usepackage{booktabs}
\usepackage[dvipsnames]{xcolor}
\usepackage[breaklinks,colorlinks=true,citecolor=blue,linkcolor=magenta,urlcolor=blue]{hyperref}
\usepackage{numdef}
\usepackage{caption}
\usepackage{subfigure}
\usepackage{rotating}
\DeclareCaptionLabelFormat{continued}{#1~#2 (continued)}
\newcommand{\msun}{${\rm M}_{\odot}$}
\newcommand{\rsun}{${\rm R}_{\odot}$}
\newcommand{\lsun}{${\rm L}_{\odot}$}
\newcommand{\teff}{$T_\mathrm{eff}$}
\newcommand{\logg}{$\log g$}
\newcommand{\logy}{$\log n(\mathrm{He})/n(\mathrm{H})$}

\num\def\BD+25{BD+25$^{\circ}\,$4655}

\num\def\A9{\textsc{Atlas}{\footnotesize9}}
\num\def\A12{\textsc{Atlas}{\footnotesize12}}
\newcommand{\kms}{km\,s$^{-1}$}

\newcommand{\kmsec}{\,\mbox{$\mbox{km}\,\mbox{s}^{-1}$}}

\num\def\J041{J0415+2528}
\num\def\J080{J0809$-$2627}
\num\def\J130{J1303+2646}
\num\def\J160{J1603+3412}
\num\def\J123{J1233$-$6749}
\num\def\J125{J1256$-$5753}
\num\def\J144{J1444$-$6744}
\num\def\J071{J0714$-$2245}  
\num\def\J134{J1346$-$4025}  
\num\def\J194{J1946$-$4756}  

\begin{document}

\title{Discovery of three magnetic helium-rich hot subdwarfs with SALT}

\author{M.\ Dorsch\inst{1}, C.\ S.\ Jeffery\inst{2}, A.\ Philip Monai\inst{2,3}, C.\ A.\ Tout\inst{4},  E.\ J.\ Snowdon\inst{2,3}, I.\ Monageng\inst{5,6}, L.\ J.\ A.\ Scott\inst{2}, B.\ Miszalski\inst{7}, V.\ M.\ Woolf\inst{8}  }  

\institute{
Institut für Physik und Astronomie, Universität Potsdam, Haus 28, Karl-Liebknecht-Str.\ 24/25, 14476 Potsdam-Golm, Germany
\email{dorsch@uni-potsdam.de} 
\and
Armagh Observatory and Planetarium, College Hill, Armagh BT61 9DG, Northern Ireland
\and
School of Mathematics and Physics, Queen's University Belfast, Belfast, BT7 1NN, UK
\and
Institute of Astronomy, The Observatories, Madingley Road, Cambridge CB3 OHA, UK 
\and
South African Astronomical Observatory, P.O. Box 9, Observatory Rd., Observatory 7935, Cape Town, South Africa
\and
Department of Astronomy, University of Cape Town, Private Bag X3, Rondebosch 7701, South Africa
\and
Australian Astronomical Optics, Faculty of Science and Engineering, Macquarie University, North Ryde, NSW 2113, Australia
\and
Department of Physics, University of Nebraska at Omaha, 6001 Dodge St, Omaha, NE 68182-0266, USA
}

\date{Received ; accepted }

\abstract
{

\noindent Magnetic fields with strengths ranging from 300 to 500\,kG have recently been discovered in a group of four extremely similar helium-rich hot subdwarf (He-sdO) stars. 
In addition to their strong magnetic fields, these He-sdO stars are characterised by common atmospheric parameters, clustering around \teff\ = 46\,500\,K, a $\log g/\mathrm{cm}\,\mathrm{s}^{-1}$ close to 6, and intermediate helium abundances. 
Here we present the discovery of three additional magnetic hot subdwarfs, J123359.44$-$674929.11, J125611.42$-$575333.45, and
J144405.79$-$674400.93. 
These stars are again almost identical in terms of atmospheric parameters, but, at $B \approx$ 200\,kG, their magnetic fields are somewhat weaker than those previously known. 
The close similarity of all known He-sdOs implies a finely tuned formation channel. We propose the merging of a He white dwarf with a H+He white dwarf. 
A differential rotation at the merger interface may initiate a toroidal magnetic field that evolves via a magnetic dynamo to produce a poloidal field. This field is either directly visible at the surface or might diffuse towards the surface if initially buried. 
We further discuss a broad absorption line centred at about 4630\,\AA\ that is common to all magnetic He-sdOs. This feature may not be related to the magnetic field but instead to the intermediate helium abundances in these He-sdO stars, allowing the strong \ion{He}{ii} 4686\,\AA\ line to be perturbed by collisions with hydrogen atoms. 
}


\authorrunning{Dorsch et al.}
\titlerunning{Discovery of three magnetic He-sdOs with SALT}

\maketitle

\section{Introduction}

Hot subdwarf stars of spectral type O and B (sdO/B) are low-mass, typically helium-burning stars that have very thin or no hydrogen envelopes \citep[for reviews, see][]{heber09,heber16}. 
Most of these stars are thought to be formed by binary evolution, including stellar merges in the case of single hot subdwarfs \citep{Han2002}. 
Helium-rich hot subdwarfs (He-sdOs) are the natural outcome of double helium white dwarf (WD) mergers \citep{Saio2000}. 
Recently, four He-sdOs were discovered to host magnetic fields  ranging from $B=300$ to 500\,kG \citep{Dorsch2022, Pelisoli2022}. 
These stars form a small but homogeneous class that can be called iHe-sdOH, in analogy to the magnetic DAH WD stars \citep[][]{Bagnulo2021}. 
All known magnetic He-sdOs have effective temperatures $T_\mathrm{eff} \approx 46\,500$\,K, surface gravities  slightly below $\log g/\mathrm{cm}\,\mathrm{s}^{-1}$ = 6\footnote{Surface gravities are stated in cgs units throughout this paper. }, and a helium abundance of $\log n(\mathrm{He})/n(\mathrm{H})\approx 0$, which is unusually low for He-sdO stars. 
It seems plausible that their magnetic fields were generated during their formation in a WD merger event \citep{Garcia-Berro2012}. 
However, it remains unclear why the vast majority of He-sdOs do not show magnetic fields, given that they are also thought to be formed in WD mergers. 

Here we present the discovery of three additional magnetic He-sdOs, J123359.44$-$674929.11, J125611.42$-$575333.45, and J144405.79$-$674400.93, hereinafter \J123, \J125, and \J144, respectively\footnote{\J123\ $\equiv$ \textit{Gaia} DR3 5856712448201848320, \J125\ $\equiv$ \textit{Gaia} DR3 6060271804721212672, and \J144\ $\equiv$ \textit{Gaia} DR3 5848132374839987840. }. These stars closely mirror the atmospheric parameters of the known magnetic He-sdOs, but their average field strengths are slightly lower, between 180 and 220\,kG. 
We also provide a list of six additional He-sdO stars that show a broad absorption feature at about 4630\,\AA\ -- a feature that is also observed in all magnetic He-sdOs \citep{Pelisoli2022}. A potential non-magnetic origin of this spectral feature is briefly discussed.  

\section{Spectroscopic observations}
\label{sect:obs}

All stars discussed here were observed as part of the ongoing Southern African Large Telescope (SALT) survey for He-sdO stars. This survey is described in detail by \cite{Jeffery2021}. 
Spectra were taken with the Robert Stobie Spectrograph (RSS) with a slit width of 1$\arcsec$ and the PG2300 grating at grating angles of both 30.5$^\circ$ and 32$^\circ$. 
This results in a spectral resolution of $\Delta \lambda \approx 1.05$\,\AA\ and coverage of the 3850\,\AA\ to 5090\,\AA\ range without gaps. 
We searched for magnetic fields in all 592 stars observed so far in this SALT/RSS programme. 

To check for radial velocity and magnetic field variability, we obtained multiple SALT/RSS spectra of the new magnetic He-sdOs, \J123, \J125,\ and \J144.  
For each of these stars, the first (discovery) spectrum was taken in May 2023, and three follow-up spectra were taken at the end of January 2024, spaced by several days. 
A summary of these observations is given in Table \ref{tab:spec_mag}.  

  \begin{figure}
        \centering
        \includegraphics[width=0.99\columnwidth]{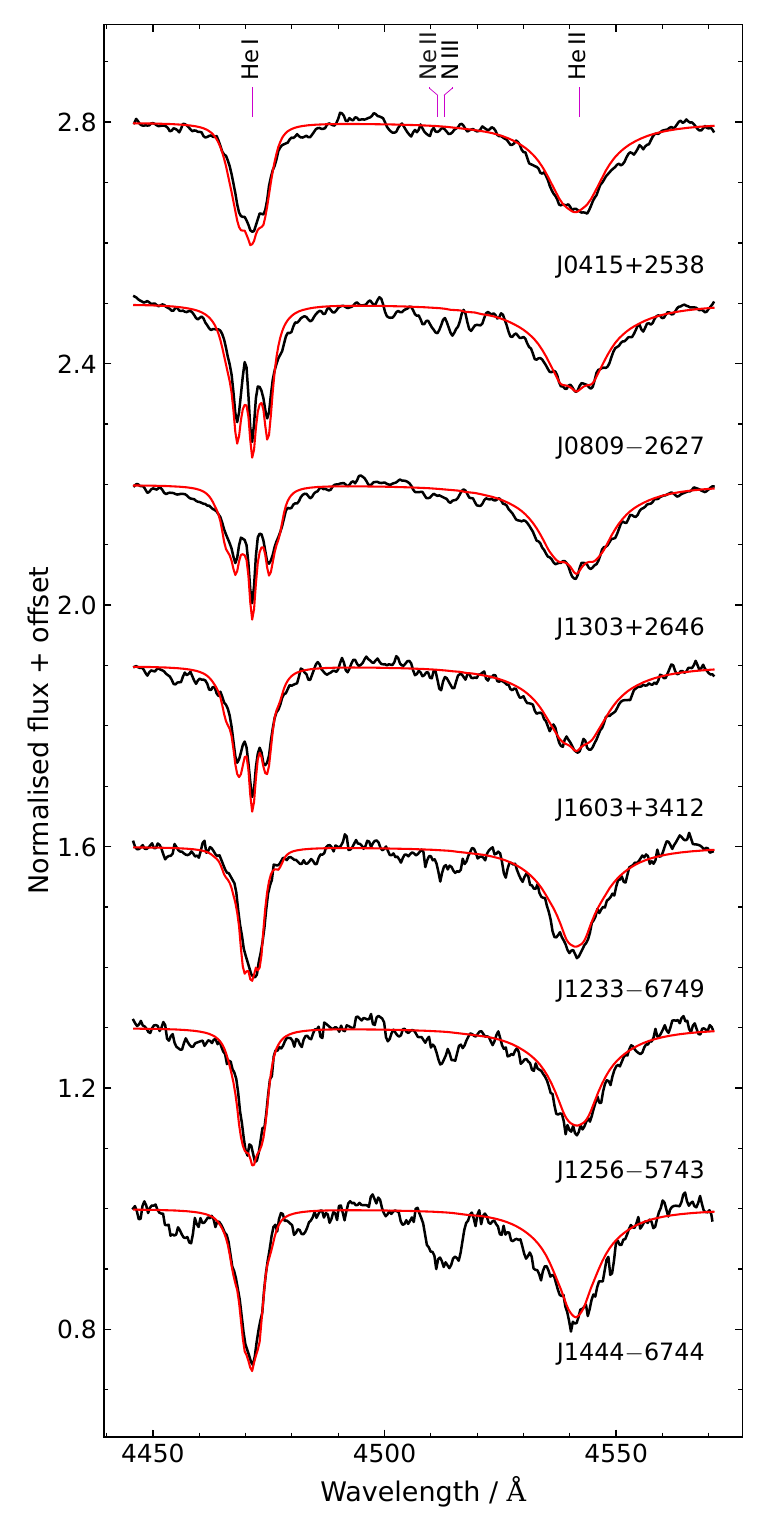}
    \vspace{-10pt}
        \caption{\ion{He}{i} 4471\,\AA\ and \ion{He}{ii} 4542\,\AA\ for magnetic He-sdOs: previously known stars (top) and new ones from SALT/RSS (bottom). Observations are black and best-fit models red. All spectra are at the rest wavelength, offset in steps of 0.3 in normalised flux.
    They were convolved to the RSS resolution (1.05\,$\mathrm{\AA}$) except for that of J0415+2538, which has a worse resolution (1.9\,$\mathrm{\AA}$). 
 }
        \label{fig:tree_heI}
 \end{figure}
 \begin{figure}
        \centering
        \includegraphics[width=0.99\columnwidth]{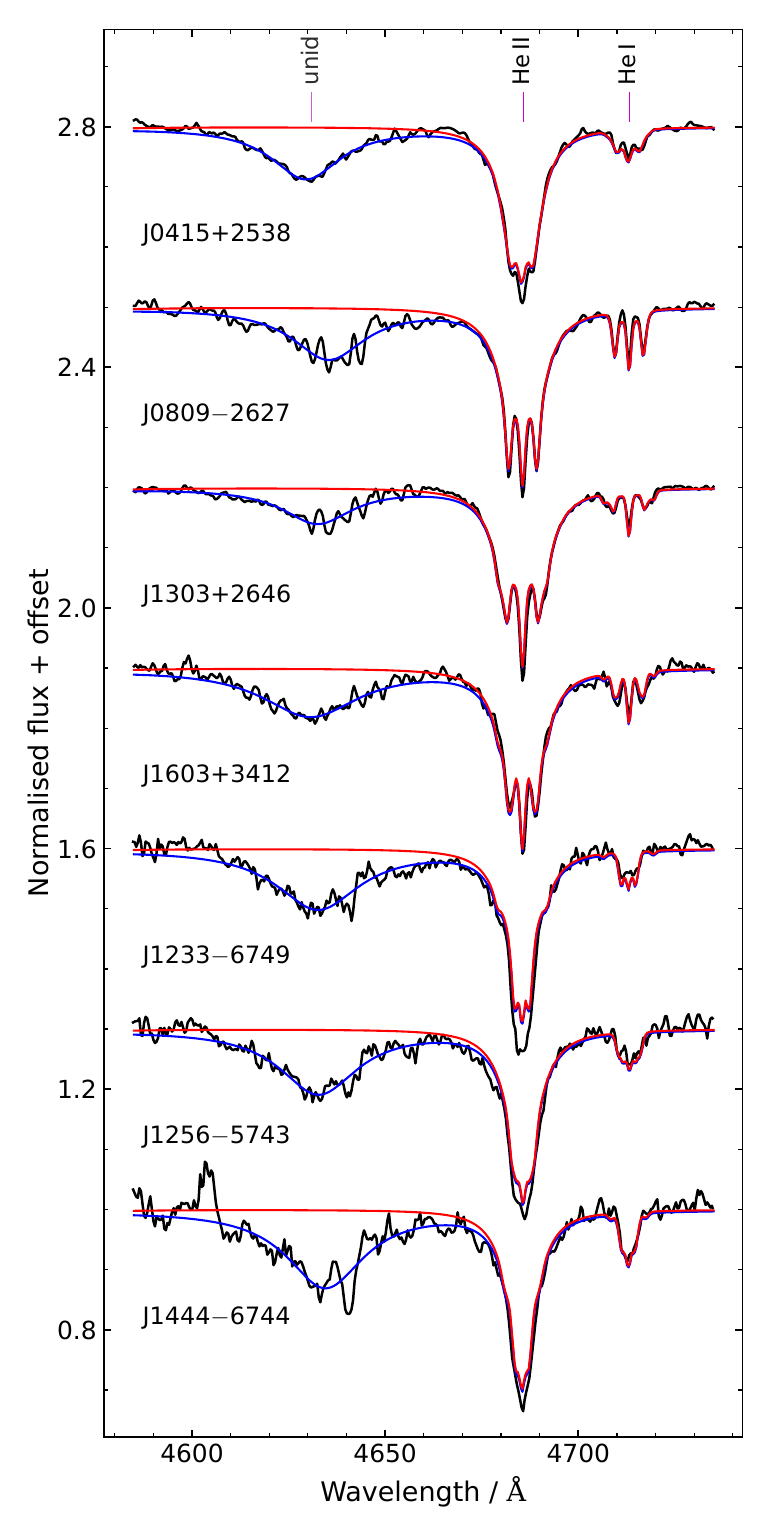}
  \vspace{-10pt}
        \caption{Like Fig.\ \ref{fig:tree_heI}, but for the unidentified feature, \ion{He}{ii} 4686\,\AA,\ and \ion{He}{i} 4713\,\AA. Lorentz profile fits to the unidentified feature at about 4630\,\AA\ are shown in blue. 
 }
        \label{fig:tree_heII}
 \end{figure}
\begin{table*}
\setstretch{1.05}
\caption{Spectroscopic atmospheric parameters for confirmed magnetic He-sdOs (upper), new magnetic stars (centre), and He-sdOs showing the 4630\,\AA\ feature (lower). 
}
\label{tab:specfit}
\vspace{-12pt}
\begin{center}
\begin{tabular}{l c c c c c c}
\toprule
\toprule
Star  & \teff\ / K  & $\log g/\mathrm{cm}\,\mathrm{s}^{-1}$   & \logy\   & $B_\mathrm{av}$ / kG &  $v_\mathrm{rad}$ / \kms & S/N \\ 
\midrule
\J041$^\ast$       &   $46580\pm 1500$     & $5.98\pm 0.25$  &  $-0.10\pm 0.15$ &  $305 \pm 20$  & $-17\pm 10$ & -- \\
\J080$^\dagger$       &   $44900\pm 1000$     & $5.93\pm 0.15$  &  $\phantom{+}0.28\pm 0.10$ &  $353 \pm 10$ & $\phantom{-}33\pm 2$ & -- \\
\J130$^\ast$      &   $47950\pm 1500$     & $5.97\pm 0.30$  &  $\phantom{+}0.25\pm 0.15$ &  $450 \pm 20$  & $-37\pm 8$ & --\\
\J160$^\ast$       &   $46450\pm 1500$     & $6.06\pm 0.20$  &  $\phantom{+}0.07\pm 0.15$ &  $335 \pm 15$  & $\phantom{-0}6\pm 5$ & -- \\
\midrule
\J123\       &   $47680\pm 1000$     & $5.60\pm 0.13$  &  $\phantom{+}0.11\pm 0.05$ & $222 \pm 10$ & $\phantom{-}82\pm 15$ & 110 \\
\J125\       &   $46980\pm 1000$     & $5.75\pm 0.16$  &  $\phantom{+}0.00\pm 0.05$ &  $216 \pm 24$  & $\phantom{-0}5\pm 12$ & 110 \\
\J144\       &   $45990\pm 1000$     & $5.55\pm 0.15$  &  $\phantom{+}0.17\pm 0.06$ &  $184 \pm 14$    & $\phantom{-0}1\pm 10$ & \phantom{0}90\\    
\midrule
EC20577$-$5504 &  $46610 \pm 1000$ & $6.09 \pm 0.12$ & $\phantom{+}0.19 \pm 0.05$ & $\leq \phantom{0}59 \pm \phantom{0}3$   & $-42\pm 10$ & 180 \\
EC22332$-$6837 &  $46990 \pm 1000 $ & $5.93 \pm 0.12$  & $\phantom{+}0.28  \pm 0.06$  &  $  \leq \phantom{0}82 \pm \phantom{0}5$      & $-101\pm 10$ & 100 \\
PG1625$-$034   &  $46570 \pm 1100 $ & $5.67 \pm 0.16$  & $\phantom{+}0.49  \pm 0.09 $  &  $\leq 128 \pm \phantom{0}9$               & $-84 \pm 11$ & \phantom{0}40\\
\J071\       &  $44400 \pm 1000$ & $5.61\pm 0.13$ & $-0.26 \pm 0.05$ & $\leq 107 \pm \phantom{0}4$     & $\phantom{+}33\pm 10$ & \phantom{0}70 \\
\J134\       &  $45070 \pm 1000$ & $5.83\pm 0.12$ & $\phantom{+}0.13 \pm 0.05$ & $\leq \phantom{0}68 \pm \phantom{0}4$   & $-17\pm 10$ & 130 \\
\J194\       &  $46730 \pm 1100 $ & $5.84  \pm 0.16$  & $\phantom{+}0.12  \pm 0.08$  &  $\leq 129 \pm \phantom{0}9$                & $-41\pm 11$ & \phantom{0}40 \\
\bottomrule 
\end{tabular}
\end{center}
\vspace{-12pt}
\tablefoot{
$^\ast$ from \cite{Pelisoli2022}. 
$^\dagger$ from \cite{Dorsch2022}. 
\J071\ $\equiv$ \textit{Gaia} DR3 2927944940172059648, \J134\ $\equiv$ UCAC4 248-062759, and \J194\ $\equiv$ GALEX J194618.8-475617.
The last column lists the mean signal-to-noise ratio (S/N) of the RSS spectra. For \J123, \J125, and \J144, the four available spectra were co-added to obtain average magnetic field strengths, while \teff, \logg, and \logy\ were derived from the individual spectra. Estimated systematical uncertainties are stated for one sigma confidence. 
}
\end{table*}

\section{Spectral analysis}
\label{sect:atm}

A direct way to identify a magnetic hot subdwarf is to resolve line splitting caused by the linear Zeeman effect. 
Hydrogen and helium lines are split into three components if the field is strong enough and appear broadened if the field is weaker. 
In addition, all of the previously known iHe-sdOH stars show a strong feature at about 4630\,\AA. The origin of this remains unidentified \citep[][their Fig.\ 5]{Pelisoli2022}. 
We visually inspected the RSS spectra of all stars obtained as part of the SALT survey so far to identify possibly magnetic He-sdOs using both of these features. 

We then used the model grid of \cite{Dorsch2022} to perform $\chi^2$ fits\footnote{These fits were performed in the \textsc{isis} framework \citep{Houck2000} using the \cite{Irrgang2014} method.} to these spectra and thus obtain best-fit magnetic field strengths, as well as the atmospheric parameters \teff, \logg, and \logy. The best-fit parameters for all stars are listed in Table \ref{tab:specfit}. 
These models are based on \textsc{Tlusty} \citep{Hubeny2017c} model atmospheres in non-local thermodynamic equilibrium. The inclusion of the magnetic field is simplistic: it is assumed to be homogeneous across the stellar surface, and linear Zeeman splitting is only applied during the \textsc{Synspec} \citep{Hubeny2017a} spectral synthesis calculations\footnote{The details are described in Appendix B of \cite{Dorsch2022}.}. 
While these models are not a realistic approximation of the actual field geometry, they are sufficient to detect Zeeman splitting and to estimate an average field strength.

\subsection{Three new magnetic He-sdOs}
\label{sect:magn}

\begin{figure}
        \centering
        \includegraphics[width=0.99\columnwidth]{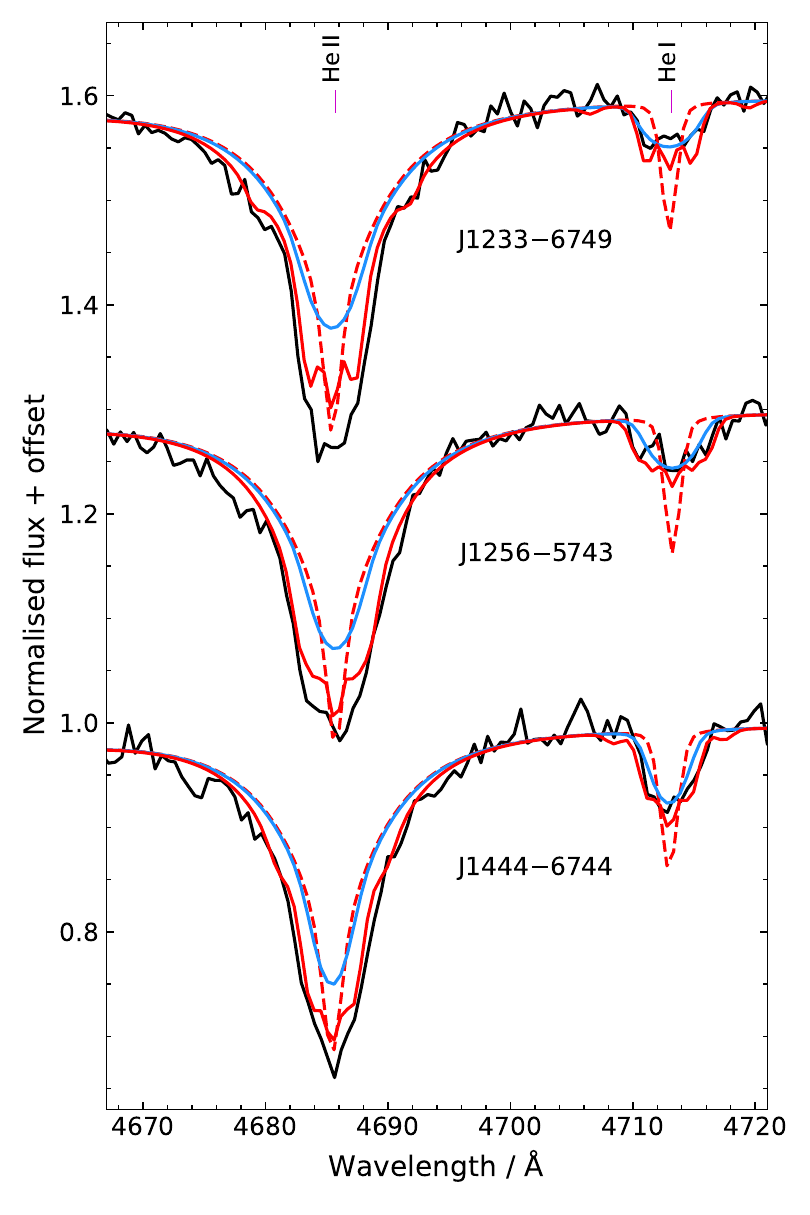}
  \vspace{-10pt}
        \caption{Detailed view of \ion{He}{ii} 4686\,\AA\ and \ion{He}{i} 4713\,\AA\ for the new magnetic He-sdOs. The  merged RSS spectra (black) are best matched by Zeeman-split models (red). Models with the best-fit atmospheric parameters but zero magnetic field (dashed red) are excluded even if $v \sin i$ is allowed to vary (blue). 
 At zero field strength, the best-fit $v \sin i$ would equal 190, 180, and 130\,\kms\ for \J123, \J125, and \J144, respectively. 
 }
        \label{fig:mag_vs_vsini}
 \end{figure}

Significant Zeeman splitting was detected in several \ion{He}{i} lines and \ion{He}{ii} 4686\,\AA\ for three stars, \J123, \J125,\ and \J144, which we consider to be confirmed as magnetic. 
Figures\ \ref{fig:tree_heI} and \ref{fig:tree_heII} show the two most important spectral regions for these stars, \ion{He}{i} 4471\,\AA\ and the range around \ion{He}{ii} 4686\,\AA, respectively.
The Zeeman splitting observed is weaker than in the four previously known magnetic He-sdOs reported by \cite{Dorsch2022} and \cite{Pelisoli2022} and is, in fact, close to the detection limit of RSS. However, there is no other effect that could explain the observed line shapes. 
The only other conceivable option would be rapid rotation at about 180\,\kms. 
As shown in Fig.\ \ref{fig:mag_vs_vsini}, magnetic models provide a much superior fit to our RSS spectra when compared to non-magnetic rotating models. 
This is because rotation, unlike Zeeman splitting, can only increase the line broadening, not the line strengths.

Like \cite{Pelisoli2022}, we performed $\chi^2$ fits using two homogeneous magnetic field components that were allowed to vary in terms of field strength and contribution, expressed as a surface ratio. 
These $\chi^2$ fits resulted in the lowest average field strengths, $B_\mathrm{av}$, in hot subdwarfs to date, $222\pm10$\,kG (\J123), $216\pm24$\,kG (\J125), and $184\pm14$\,kG (\J144). 
These average field strengths were computed by taking the weighted mean of both components $B_\mathrm{mean}$ averaged over the four spectra available for each star, as stated in Table \ref{tab:spec_mag}. 
Because our spectra were taken with one- and two-day gaps, the variation in the magnetic field strength between these four spectra may result from a variable field geometry, possibly caused by rotation combined with magnetic spots. 
Individual fits are shown in Figs.\ \ref{fig:spec_var_1} to \ref{fig:spec_var_3} for the most important spectral lines; any variability in the field strengths should be confirmed with higher-quality spectra. 

In contrast to the possibly variable fields, none of the three new magnetic He-sdOs showed signs of radial velocity variability, given the approximate 10\,\kms\ accuracy achieved by RSS. 

The magnetic fields found here are very likely non-homogeneous, and more sophisticated models would be needed to further constrain their geometry. 
Such models would also benefit from a future determination of metal abundances. 
The strong \ion{N}{iii} 4511, 4515\,\AA\ lines shown in Fig.\ \ref{fig:tree_heI} seem to suggest that the three new magnetic He-sdOs are significantly enriched in nitrogen compared to the Sun, especially \J144. 
This star also seems to be enriched in neon, as evidenced by strong \ion{Ne}{ii} 4290, 4430\,\AA\ and \ion{Ne}{iii} 4453, 4459\,\AA\ lines. 
Quantitative abundances will be measured via a more detailed analysis in the future.

\begin{figure}
        \centering
        \includegraphics[width=0.99\columnwidth]{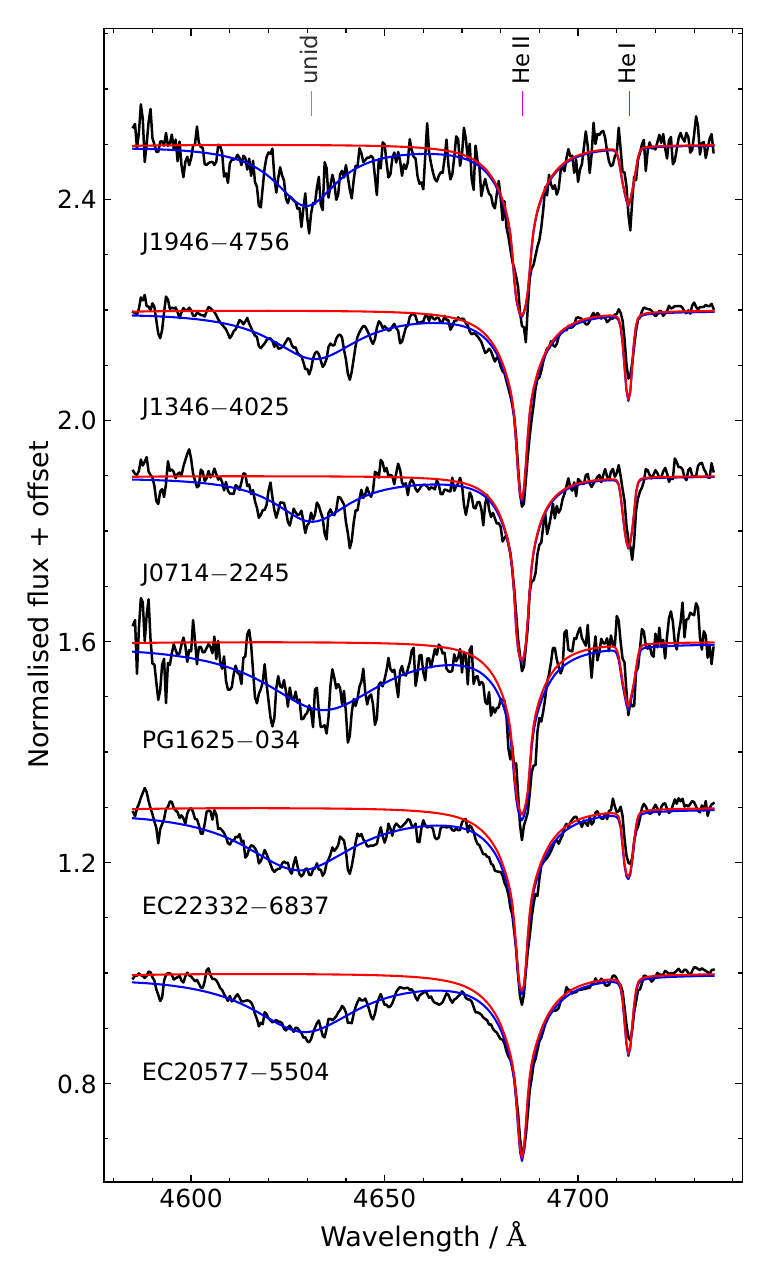}
  \vspace{-10pt}
        \caption{Like Fig.\ \ref{fig:tree_heII}, but for SALT/RSS spectra of stars that show the unidentified feature without significant Zeeman splitting. 
 }
        \label{fig:tree_feature_heII}
 \end{figure}

\subsection{The broad 4630\,\AA\ feature}
\label{sect:stars_feature}

The six other stars in our sample show no Zeeman splitting but still feature a broad absorption line at about 4630\,\AA, as shown in Fig.\ \ref{fig:tree_feature_heII}. 
This lack of Zeeman splitting provides upper limits on the mean magnetic field strength. 
The somewhat broadened helium lines of PG1625$-$034, J0714$-$2245, and J1946$-$4756 would be matched by magnetic field strengths of $128\pm 9$\,kG, $107\pm 4$\,kG, and $129\pm 9$\,kG, respectively. 
Because line broadening can be introduced by other effects, such as rotation, these stars require higher-resolution follow-up spectra to exclude or confirm the presence of a magnetic field. 
The low upper limits of $B = 59\pm 3$\,kG for EC20577$-$5504 and $B=68\pm4$\,kG for J1346$-$4025 suggest that these stars are non-magnetic. 
This would have implications for the interpretation of the 4630\,\AA\ feature, which seems to be present in all intermediately helium-rich He-sdOs, as long as $T_\mathrm{eff} \approx 46\,000$\,K and $\log g \gtrsim 5.5$.

The origin of the 4630\,\AA\ feature\footnote{It should be noted that the 4630\,\AA\ feature is in fact not always centred at a rest wavelength of exactly 4630\,\AA.
For the iHe-sdOs considered here, its centre ranges between 4629\,\AA\ and 4635\,\AA\ at Lorentzian full widths at half maximum between 20\,\AA\ and 38\,\AA.} may be completely unrelated to the magnetic field. 
As already discussed by \cite{Pelisoli2022}, this feature does not seem to be caused by Zeeman-split metal lines, diffuse interstellar bands, or even ultra-high excitation
lines, a feature that is observed for some DO-type WDs \citep{Reindl2019}. 
Instead, the 4630\,\AA\ feature seems to be related to the intermediate helium abundances: 
it could result from collisions with free hydrogen during the formation of \ion{He}{ii} 4686\,\AA. This is similar to the blue satellite lines predicted for the $\mathrm{H}\alpha$ and $\mathrm{H}\beta$ lines in WD stars by \cite{Allard2022} and \cite{Spiegelman2022}.
However, these features are  not detected here, possibly because of the much higher \teff\ and lower photospheric densities of intermediate He-sdO stars when compared to the WDs considered by \cite{Allard2022} and \cite{Spiegelman2022}. 

\section{Stellar parameters}
\label{sect:sed}

 \begin{figure}
        \centering
        \includegraphics[width=0.99\columnwidth]{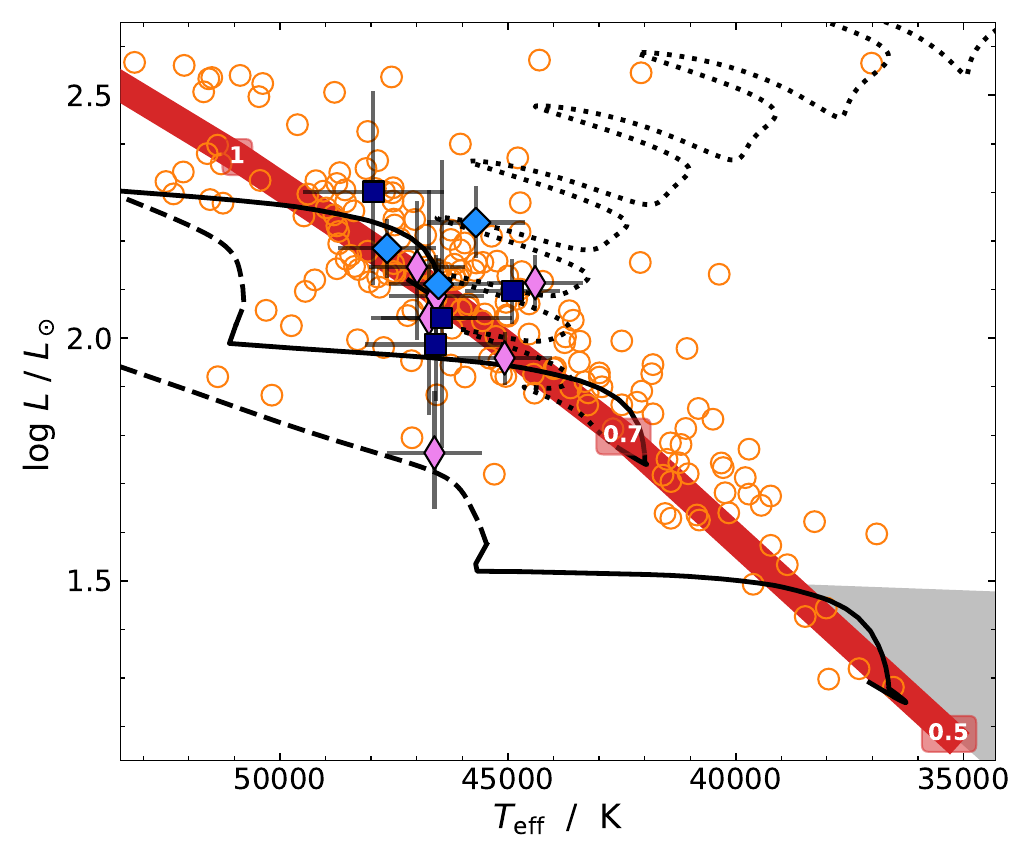}
        \caption{
Hertzsprung-Russell diagram showing known (dark blue) and new magnetic He-sdOs (light blue), as well as 4630\AA\  stars (violet). 
Merger tracks from \citet{Yu2021} for a metallicity of $Z=0.01$ and masses of 0.80, 0.65, and 0.50\,\msun\ are shown as black lines (solid for core helium-burning and dashed for helium shell-burning). The pre-helium main sequence phase is shown as a dotted line for 0.65\,\msun.
The  \citet{Paczynski1971} HeZAMS is shown in red, and the grey region represents the extreme horizontal branch.  
Orange circles represent other He-sdOs from the SALT sample. 
 }
        \label{fig:hrd}
 \end{figure}

The spectral analysis performed in Sect.\ \ref{sect:atm} provides atmospheric parameters, in particular the effective temperature and surface gravity. They can be combined with the \textit{Gaia} Data Release 3 \citep[DR3;][]{gaia2023} parallax, $\varpi$, and an angular diameter, $\Theta,$ from photometric flux measurements to derive radii $R = \Theta/(2\varpi)$ and luminosities $L = 4\pi R^2 \sigma_\mathrm{SB} T^4_\mathrm{eff}$, where $\sigma_\mathrm{SB}$ is the Stefan–Boltzmann constant. 
To this end, we performed spectral energy distribution (SED) fits to all the He-sdOs discussed here.
As usual, we used \textit{Gaia} DR3 parallaxes with  zero-point corrections applied according to \cite{Lindegren2021} and  inflated uncertainties following \cite{El-Badry2021}. 
We could also estimate stellar masses via $M = g R^2/G$. 
These masses remain poorly constrained due to the high uncertainty in the surface gravity, $g$, resulting from  additional line broadening introduced by non-homogeneous magnetic fields. 
To account for such systematic effects, we added estimated uncertainties of 1000\,K in \teff, 0.15 in \logg,\ and 0.1 in \logy\ in quadrature to the statistical uncertainties for the computation of stellar parameters. 
The resulting stellar parameters are listed in Table \ref{tab:stellar}. 

While the stellar masses are not well determined, the derived radii and luminosities are quite precise. 
All magnetic He-sdO stars are located close the zero-age helium main sequence (HeZAMS) in the Hertzsprung-Russell diagram (Fig.\ \ref{fig:hrd}).
This is not uncommon for He-sdO stars, as the comparison to the remaining SALT sample of He-sdO stars shows\footnote{A detailed analysis of the full SALT sample will be published in a separate paper.}. 
The most likely formation scenario for all (non-binary) He-sdOs is the merging of two WDs, at least one of which is a helium-core WD \citep{Han2002}. 
The clustering of the now seven confirmed magnetic He-sdO stars at about 46\,500\,K on the HeZAMS further strengthens the proposition that these stars formed through a very specific channel. 
If their masses are interpolated from their location on the \cite{Paczynski1971} HeZAMS in the Hertzsprung-Russell diagram, their mass distribution is centred at $M^\mathrm{HeZAMS} = 0.85\pm 0.05$\,\msun, which is at the upper end of what is achievable by double He-WD mergers \citep{Zhang2012}. 
This mass is a rough estimate given that (1) all magnetic He-sdOs retain significant amounts of hydrogen in their atmospheres, while the HeZAMS assumes a pure helium composition and (2) the stars may also be somewhat evolved. 
The interpolated HeZAMS masses for all stars are also listed in Table \ref{tab:stellar}. 

\section{Galactic kinematics}
\label{sect:kin}

Galactic space velocities for the confirmed magnetic He-sdOs were calculated from their radial velocity (Table \ref{tab:specfit}) and the parallax and proper motion provided by {\it Gaia} DR3. Their Galactic orbits were computed with \verb|galpy|\footnote{http://github.com/jobovy/galpy}, a python package for Galactic-dynamic calculations \citep{bovy2015}. 
The resulting kinematic parameters are summarised in Table \ref{tab:galpy}, while Fig.\ A.7 shows the projection of the computed trajectories to the vertical distance ($z$) against the Galactic radius ($R$) plane. 
The stars are roughly classified as Galactic thin disk, thick disk, or halo stars according to the  \cite{PhilipMonai2024} method. 
\J125\ and \J144\ are on the almost circular orbits typical of thin disk stars, while \J123\ is likely a thick disk star given its larger eccentricity of $e = 0.31\pm 0.01$. 
Similar calculations for the previously discovered magnetic He-sdOs show that, of the seven known magnetic stars, four are part of the thin disk, two are thick disk stars, and one ( J1303+2646) is part of the Galactic halo. 
This suggests that magnetic He-sdO stars can be formed in both young and old Galactic populations. 
This is consistent with the large range of possible delay times between the merging of double He-WD binaries and their initial formation as main sequence stars \citep{Marsh1995,Yu2011}. 

\section{Light curves}
\label{sect:lc}

One might expect rotational light curve variability for strongly magnetic stars with inhomogeneous fields. 
Magnetic fields can cause dark or bright spots on a star's surface, resulting in variations in brightness as the star rotates.
For instance, bright magnetic spots are predicted to exist in O/B/A-type main sequence stars, caused by (sub-)surface convection zones \citep{Cantiello2011,Cantiello2019} or chemical spots \citep{Shultz2018,Krticka2021}, and are observed at low amplitudes in A- and late B-type stars \citep{Balona2019}. 
Similar variability was discovered for the helium-rich early B-type star HD 144941 \citep{Jeffery2018}, a magnetic main sequence star \citep{Przybilla2021,Shultz2021}. 
Sub-surface convection zones are also predicted to exist in He-sdO stars \citep{Groth1985} due to the ionisation of \ion{He}{ii}. 
\cite{Momany2020} observed brightness variations in several hot subdwarfs in globular clusters at periods of the order of days and attributed this variability to rotational modulation caused by magnetic spots.  
\cite{Krticka2024} performed a spectroscopic analysis of three of these stars, ruling out binarity as the cause of the variations and detecting \teff\ variability, though the presence of magnetic spots remains uncertain. Finally, flux variations in several cool magnetic WDs are proposed to be caused by star spots \citep{Brinkworth2013}. 

Considering the magnetic nature of the He-sdOs discussed here, the presence of magnetic spots on these stars seems at least plausible. We therefore searched for light curves in the Transiting Exoplanet Survey Satellite \citep[TESS;][]{Ricker2014} archive up to sector 70. Unfortunately, none of the new magnetic He-sdOs have been observed with TESS so far. However, TESS data are available for three of the stars that show the 4630\,\AA\ feature. We computed frequency-amplitude spectra for these stars. 

EC20577$-$5504 was observed in TESS sectors 27 and 67. 
It seems to show a highly significant variation at a period of 0.254\,days and its first harmonic (Fig.\ \ref{fig:TESS}: top). 
However, this variation is likely caused by the limited angular resolution of TESS images; the light curve seems to be contaminated by a close-by star (\textit{Gaia} DR3 6458088855502103424). 

\J134\ was observed in TESS sectors 38 and 68 and shows an approximately sinusoidal modulation with a period of 5.4\,days (Fig.\ \ref{fig:TESS}: middle). This is commensurate with a rotational modulation and  neither confirms nor excludes the presence of a magnetic field. 

No light variation is detected for EC22332$-$6837 (TESS sectors 27, 28, 67, and 68). However, the S/N is such that the modulation seen in \J134\ would lie below the detection threshold for EC22332$-$6837 (Fig.\ \ref{fig:TESS}: bottom).

\section{Summary and discussion}
\label{sect:conclusions}

We have identified three new magnetic He-sdO stars among the sample of 592 stars observed with SALT/RSS. 
This brings the total number of magnetic hot subdwarfs to seven, which form a homogeneous class. 

\subsection{Fraction of magnetic stars}

Before considering the formation of magnetic hot subdwarfs, it makes sense to re-assess their frequency among apparently non-magnetic He-sdO stars. 
Because there are 282 He-sdO stars in the current SALT sample, the fraction of magnetic He-sdOs amongst all He-sdOs in SALT can be estimated as $1.1 \pm 0.6$ per cent. 
The seven known magnetic He-sdOs have very similar atmospheric parameters. 
It may therefore be instructive to restrict the selection of comparison stars to a 3$\sigma$ range around their mean parameters, \teff\ = $46650 \pm 2900$\,K, \logg\ = $5.83 \pm 0.56,$ and \logy\ = $+0.11 \pm 0.38$. 
Of the nine iHe-sdOs in the SALT sample that satisfy these criteria, three are magnetic and six do not show detectable Zeeman splits -- these are the stars discussed in Sect.\ \ref{sect:stars_feature}. 
This puts the fraction of magnetic stars amongst these iHe-sdOs at $33 \pm 16$ per cent. 
These fractions are lower limits because the detection threshold of Zeeman splitting in the SALT/RSS sample is about 100\,kG. 

\subsection{Magnetic field}

All magnetic He-sdOs have observed field strengths of 200 to 500\,kG and very similar surface properties, with \teff\ $ \approx  46\,500$\,K, $\log g \approx 6,$ and intermediate helium abundances at \logy\ $ \approx 0$. 
That is to say, the magnetic sdO stars are very localised compared with the overall hot subdwarf population. 
A single-star origin seems unlikely for the following reasons. 
\begin{itemize}
  \setlength\itemsep{2pt}
    \item [1)] The subdwarf masses of about 0.8\,\msun\ inferred from their locations on the helium main sequence are too high to have come from post giant branch evolution, that is, following common-envelope ejection or Roche-lobe stripping. This is because helium ignition occurs when a red-giant core reaches a mass of about 0.5\,\msun,\ and the envelope must be removed close to this point if the star is to become a hot subdwarf. 
    \item [2)] Any pre-existing surface magnetic field would have been destroyed by the surface convection zone whilst the star was on the giant branch.
    \item [3)] Any pre-existing core magnetic field in a hot subdwarf would likely be destroyed by shell and core convection following helium ignition, as pointed out by \citet{Cantiello2016}. 
    While less convection would result from a non-degenerate He ignition, this cannot lead to the formation of a single hot subdwarf star, given that there is no way to remove the hydrogen envelope. 
    \item [4)] Differential rotation induced by the contraction of the residual star from the giant branch to become a subdwarf could lead to a dynamo-driven field. This raises the question of why this effect would not have occurred in all hot subdwarfs. There is no evidence for any detectable field in any other hot subdwarfs \citep{Landstreet2012,Mathys2012, Randall2015}.
\end{itemize}

A double-star origin seems possible. The most promising scenarios include a double He-WD merger or a merger of a hybrid HeCO-WD with a He-WD \citep{Justham2011}, as already discussed by \cite{Dorsch2022} and \cite{Pelisoli2022}. 
Such mergers would explain their helium-enriched surfaces and apparently single nature.  
This origin may be able to generate the magnetic fields but we are still left with the same question: why do not all massive subdwarfs have strong fields, in particular other He-sdO stars? 
A finely tuned formation scenario seems to be required.

Double-degenerate merger routes involving a fossil field in either progenitor were also considered. 
\begin{itemize}
  \setlength\itemsep{2pt}
    \item [1)] A fossil field in the accretor fails because off-centre helium ignition creates convective shells after every shell flash, extending from close to surface, after the first shell flash, through to the core, once core-burning is fully established. Flash-driven convection would progressively dissipate the field throughout the interior.  
    \item [2)] A fossil field in the donor fails because the donor material is fully mixed and/or fully convective at or near the merger.
Hence, a fossil field cannot survive a double WD merger leading to the creation of a helium-burning subdwarf.
\end{itemize}

This leaves the merging of two low-mass WDs as the only realistic formation scenario for magnetic He-sdOs. 
Recently, \cite{Pakmor2024} conducted magnetohydrodynamic simulations of the merging of two He-WDs with masses of either 0.3 + 0.25\,\msun\ or 0.3 + 0.3\,\msun. They find that strong magnetic fields can be generated during both mergers. Their simulations indicate that if helium ignites off-centre, strong convection will erase these fields. However, if helium ignites in the core, the fields can persist. This core ignition scenario is likely when the merging WDs have similar core densities, which corresponds to nearly equal masses. While the observed magnetic He-sdOs appear to be more massive than those in the simulations, the underlying physics might be similar. Future research should use the results from these magnetohydrodynamic simulations to develop post-merger evolutionary models, which can predict the properties of magnetic He-sdOs once they settle onto the helium main sequence.

Because the helium enrichment in magnetic He-sdOs is less extreme than in the more common non-magnetic He-sdOs, new models that include a treatment of the magnetic field need to be able to retain a significant amount of surface hydrogen. 
Our proposal is that the magnetic subdwarfs arise from the merging of a He-WD with a H+He-WD.  
As they merge, the H+He-WD is destroyed and fully mixed. Most of its mass condenses onto the He-WD's surface. 
Differential rotation (shear) exists at the interface between the accretor and the accreted material. 
This initiates a toroidal magnetic field. 
This toroidal field evolves via a magnetic dynamo to produce a poloidal field, which is then visible at the surface. 

A suitable double-WD progenitor would have a combined mass about 0.8\,\msun, with component masses of about 0.4 to 0.5\,\msun\ for the He-WD and about 0.3 to 0.4\,\msun\ for the H+He-WD. 
The H+He-WD could be marginally the more massive of the pair, so long as it has the larger radius. 
The hydrogen content of either component cannot be too large because the location of iHe-sdOH stars on the helium main sequence requires that there be negligible hydrogen beneath the atmosphere. 
Potential systems have been identified theoretically \citep{Han1998} and observed in both the Galactic disk and halo \citep{Brown2020}. 
They are believed to arise as a consequence of two common-envelope ejection episodes in an intermediate-mass binary, where both episodes occur when each donor star is on the red-giant branch. 
The condition that some hydrogen remains on the surface of either He-WD is that the common-envelope component shrinks more quickly than the orbit decays, and before all of the hydrogen is completely removed. 
\citet{Hall2016} discuss similar cases for a combined mass of about 0.5\,\msun\ and find that very little hydrogen could survive. 
An important feature of the merging double WD model is that the first helium-shell flash drives a strong convection zone upwards. 
If this zone reaches the surface, it reveals carbon and removes residual hydrogen \citep{Zhang2012}. 
Such a convective pulse would also redistribute angular momentum from the original shear layer, moving the shear layer towards the surface. 
The absence of carbon in the spectra of magnetic He-sdOs implies that complete mixing has not occurred and that a field-generating shear layer could still exist beneath the stellar surface.  
Such constraints may contribute in part to the fine-tuning  required to model the iHe-sdOH stars.

This proposal needs to be tested more fully in a number of ways. We need to: 
\begin{itemize}
\setlength\itemsep{2pt}\item [1)] Demonstrate that the merger leads to sufficient differential rotation in the mixed envelope to give rise to a strong magnetic field.  
\item [2)] Determine  properties, such as the mass, residual hydrogen, and differential rotation, necessary for the merged system to generate a dynamo via population synthesis or other means. 
\item [3)] Establish the progenitor properties at the merge necessary to yield (2), such as the He-WD mass, H+He-WD mass, and H+He-WD core mass. 
\end{itemize}
All of these issues likely need to be resolved if we are to understand why there is a small sweet spot for generating strong magnetic fields in hot subdwarfs. The proposed merging of a He-WD with a H+He-WD by itself cannot explain why the observed magnetic He-sdOs are so similar. 
However, \cite{Pakmor2024} suggest that only WDs with similar core densities can maintain a strong surface magnetic field in the He-sdO remnant. This finding may explain the observed clustering in the Hertzsprung-Russell diagram because similar core densities in the merging WDs would result in similar core masses for the He-sdOs. 

Unfortunately, the detailed surface metal abundances of the magnetic He-sdOs discovered here are not known, because of the limited quality of the available spectra. 
Far-UV spectra taken with the \textit{Hubble} Space Telescope could solve this issue but are presently not available.

\subsection{Kinematics and light curves}

We further performed Galactic orbit calculations for all seven known magnetic He-sdO stars. 
Given that we find four stars in the thin disk, two in the thick disk, and one in the Galactic halo, magnetic He-sdOs seem to exist in all Galactic populations, both old and young. 
Unfortunately, no TESS light curves are available for the three new He-sdOHs. 
Light curve variability therefore remains undetected for all seven He-sdOHs, and follow-up time series photometry should be performed.

\subsection{4630\AA\ absorption}

A broad absorption feature at 4630\,\AA\ was identified in six additional intermediate He-sdO stars. 
These six stars are not necessarily magnetic, given that they do not show Zeeman splitting. 
Even though the 4630\,\AA\ feature is present in all magnetic He-sdOs \citep{Pelisoli2022}, it seems to be unrelated to the magnetic field. 
Instead, it might result from collisional perturbations to the \ion{He}{ii} 4686\,\AA\ line caused by ionised hydrogen, a process that was studied in detail by \cite{Allard2022} and \cite{Spiegelman2022} for the case of the H$\alpha$ and  H$\beta$ lines. 

\begin{acknowledgements}
We thank N.\ F.\ Allard for her useful comments on the 4630\,\AA\ feature and the anonymous referee for their detailed comments. 
MD is supported by the Deutsches Zentrum für Luft- und Raumfahrt (DLR) through grant 50-OR-2304. 
CAT thanks Churchill College for his fellowship. 
This work has made use of data from the European Space Agency (ESA) mission {\it Gaia} (\url{https://www.cosmos.esa.int/gaia}), processed by the {\it Gaia} Data Processing and Analysis Consortium (DPAC, \url{https://www.cosmos.esa.int/web/gaia/dpac/consortium}). Funding for the DPAC has been provided by national institutions, in particular the institutions participating in the {\it Gaia} Multilateral Agreement.
This paper includes data collected by the TESS mission, which are publicly available from the Mikulski Archive for Space Telescopes (MAST).
Funding for the TESS mission is provided by NASA’s Science Mission directorate. This research has made use of NASA's Astrophysics Data System.

\end{acknowledgements}

\bibliographystyle{aa}
\bibliography{msdO_SALT.bib}

\begin{appendix}

\begin{onecolumn}

\section{Additional material}
\label{appendix}

At the editor's request, Figs. A.1 to A.8 are available on Zenodo only.

\begin{table}[!h]
\caption{
RSS spectra of the three new magnetic He-sdOs. 
}\label{tab:spec_mag}
\vspace{-14pt}
\setstretch{1.08}
\begin{center}
\begin{tabular}{ccccccccc}
\toprule
\toprule
Star & JD & $t_\mathrm{exp}/\mathrm{s}$ & S/N & $B_1/\mathrm{kG}$ & $B_2/\mathrm{kG}$ & $A_2/A_1$ & $B_\mathrm{mean}/\mathrm{kG}$ & ID \\
\midrule
\J123 &   2460069.27553 & 600 & 81 & $178\pm 4$ & $625\pm 26$ & $0.14\pm 0.02$ & $232\pm \phantom{0}6$ & a  \\
  &   2460337.55751 & 600 &  43 & $178\pm 8$ & $550\pm 50$ & $0.14\pm 0.04$ & $225\pm 11$ & b \\
  &   2460340.59014 & 600 &  95 & $178\pm 4$ &$568\pm 30$ & $0.15\pm 0.02$ & $227\pm \phantom{0}7$ & c \\
  &   2460341.50690 & 600 & 35 & $165^{+70}_{-30}$ & $250^{+50}_{-70}$ & $0.87^{+13}_{-0.7}$ & $204\pm 42$ & d\\[1mm]
\J125 &  2460069.27480 & 500 &  31 & $260^{+25}_{-170}$ & $130^{+30}_{-40}$ & $0.77^{+0.61}_{-0.46}$ & $201\pm44$ & a\\
  &   2460334.54060 & 500 &  67 & $164 \pm 8$ & $318 \pm 16$ & $0.43 \pm 0.10$ & $211\pm 12$ & b\\
  &  2460336.52242 & 500 &  78 & $140 \pm 8$ & $270 \pm 10$ & $0.76 \pm 0.13$ & $196\pm \phantom{0}8$& c \\
  &   2460341.49358 & 500 &  60 & $212 \pm 4$ & $625 \pm 50$ & $0.12 \pm 0.02$ & $258\pm \phantom{0}8$& d\\[1mm]
\J144 &   2460071.40057 & 500 & 48 & $147 \pm 10$ & $277 \pm 13$ & $0.73 \pm 0.16$ & $202\pm \phantom{0}9$ & a\\
  &   2460364.52464 & 500 &  35 & $143 \pm 13$ & $310 \pm 37$ & $0.25 \pm 0.09$ & $189\pm 14$ & b\\
  &   2460369.51603 & 500 &  45 & $133 \pm 15$ & $248 \pm 45$ & $0.38 \pm 0.29$ & $165\pm 25$ & c\\
  &   2460371.57299 & 500 &  67 & $145 \pm 6$ & $295 \pm 16$ & $0.31 \pm 0.05$ & $180\pm  \phantom{0}7$ & d\\
\bottomrule
\end{tabular}
\tablefoot{
The last column lists the weighted average field strength $B_\mathrm{mean}$ of the two-component best-fit for each spectrum, as discussed in Sect.\ \ref{sect:magn}. $A_2/A_1$ is the surface ratio of the second field component ($B_2$) relative to the first ($B_1$). 
}
\end{center}
\end{table}

\begin{table}[!h]
\centering
\caption{
Best stellar parameters based on the combination of spectroscopic atmospheric parameters from Table \ref{tab:specfit}, the \textit{Gaia} parallax, and the SED's angular diameter. 
}
\label{tab:stellar}
\renewcommand{\arraystretch}{1.35}
\begin{tabular}{lccccccc}
\toprule
\toprule
Star & $\log \Theta / \mathrm{rad} $ & $E(44$$-$$55) / \mathrm{mag}$ & $\varpi / \mathrm{mas}$  & $R$ / \rsun  & $L$ / \lsun  & $M^\mathrm{spec}$ / \msun & $M^\mathrm{HeZAMS}$ / \msun \\
     \midrule
\J041$^\ast$   & $-11.476 \pm 0.006$ & $0.298 \pm 0.004$ & $0.49 \pm 0.06$ & $0.151^{+0.020}_{-0.016}$  & $\phantom{0}97^{+31}_{-22}$ & $0.80^{+0.60}_{-0.40}$ & $0.82 \pm 0.06$\\
\J080$^\dagger$   & $-11.246 \pm 0.005$ & $0.072 \pm 0.003$ & $0.68 \pm 0.04$ & $0.185^{+0.011}_{-0.010}$ & $125^{+19}_{-17}$ & $1.06^{+0.31}_{-0.24}$ & $0.82 \pm 0.03$\\
\J130$^\ast$   & $-11.517 \pm 0.006$ & $0.005 \pm 0.003$ & $0.33 \pm 0.07$ & $0.210^{+0.050}_{-0.040}$ & $200^{+120}_{-70}$ & $1.40^{+2.40}_{-0.70}$ & $0.90 \pm 0.07$\\
\J160$^\ast$   & $-11.806 \pm 0.006$ & $0.027 \pm 0.005$ & $0.22 \pm 0.07$ & $0.160^{+0.070}_{-0.040}$ & $110^{+120}_{-50}$ &$1.00^{+2.20}_{-0.50}$ & $0.83 \pm 0.06$\\
\midrule
\J123\        & $-11.270\pm0.004$ & $0.21\pm0.01$ &$0.66\pm0.04$ & $0.181^{+0.009}_{-0.009}$ & $153^{+21}_{-19}$ &$0.48^{+0.22}_{-0.16}$ & $0.88 \pm 0.04$\\
\J125\        & $-11.190 \pm 0.005$ & $0.25 \pm 0.01$ & $0.82 \pm 0.03$ & $0.175^{+0.007}_{-0.007}$ & $129^{+16}_{-14}$ & $0.74^{+0.34}_{-0.24}$ & $0.85 \pm 0.03$ \\
\J144\        & $-11.141\pm0.005$ & $0.36\pm0.01$ &$0.77\pm0.05$ & $0.210^{+0.015}_{-0.013}$ & $173^{+31}_{-25}$ &$0.56^{+0.27}_{-0.19}$ & $0.86 \pm 0.04$ \\
\midrule
\midrule
EC20577$-$5504  & $-11.542\pm0.004$ & $0.03\pm0.01$ &$0.54\pm0.07$ & $0.117^{+0.017}_{-0.013}$ & $\phantom{0}58^{+19}_{-13}$ &$0.62^{+0.36}_{-0.23}$ & $0.80 \pm 0.04$ \\
EC22332$-$6837  & $-11.584\pm0.005$ & $0.04\pm0.01$ &$0.33\pm0.05$ & $0.177^{+0.028}_{-0.022}$ & $140^{+50}_{-40}$ &$1.00^{+0.60}_{-0.40}$ & $0.86 \pm 0.04$\\
PG1625$-$034    & $-11.347\pm0.003$ & $0.23\pm0.01$ &$0.59\pm0.05$ & $0.168^{+0.014}_{-0.012}$ & $122^{+25}_{-20}$ & $0.54^{+0.27}_{-0.18}$ & $0.84 \pm 0.04$\\
\J071\        & $-11.180\pm0.005$ & $0.15 \pm 0.01$ & $0.76 \pm 0.03$ & $0.192^{+0.008}_{-0.008}$ & $130^{+17}_{-15}$ &$0.48^{+0.24}_{-0.16}$ & $0.82 \pm 0.03$\\
\J134\        & $-11.000\pm0.005$ & $0.07\pm0.01$ &$1.43\pm0.05$ & $0.156^{+0.006}_{-0.006}$ & $\phantom{0}91^{+11}_{-10}$ &$0.58^{+0.27}_{-0.18}$ & $^\ddagger$\\ 
\J194\        & $-11.690\pm0.005$ & $0.04\pm0.01$ &$0.29\pm0.08$ & $0.160^{+0.060}_{-0.040}$ & $110^{+90}_{-40}$ &$0.61^{+0.60}_{-0.28}$ & $0.84 \pm 0.05$\\
\bottomrule
\end{tabular}
\tablefoot{
$^\ast$ Originally from \cite{Pelisoli2022}. 
$^\dagger$ Originally from \cite{Dorsch2022}. 
$^\ddagger$ Too far away from the helium main sequence for a sensible mass estimate. 
Estimated systematic uncertainties of 1000\,K in \teff, 0.15 in \logg, and 0.1 in \logy\ were added in quadrature  to the statistical uncertainties for the computation of the radii $R$, luminosities $L$, and masses $M$. 
Stellar parameters are stated as median values with one sigma uncertainties. 
The last column states masses obtained by  interpolation in the helium main sequence of \cite{Paczynski1971}, given the uncertainties on \teff\ and $L$. 
}
\end{table}

 \begin{figure}[!h]
        \centering
        \includegraphics[width=24.2cm,angle=90]{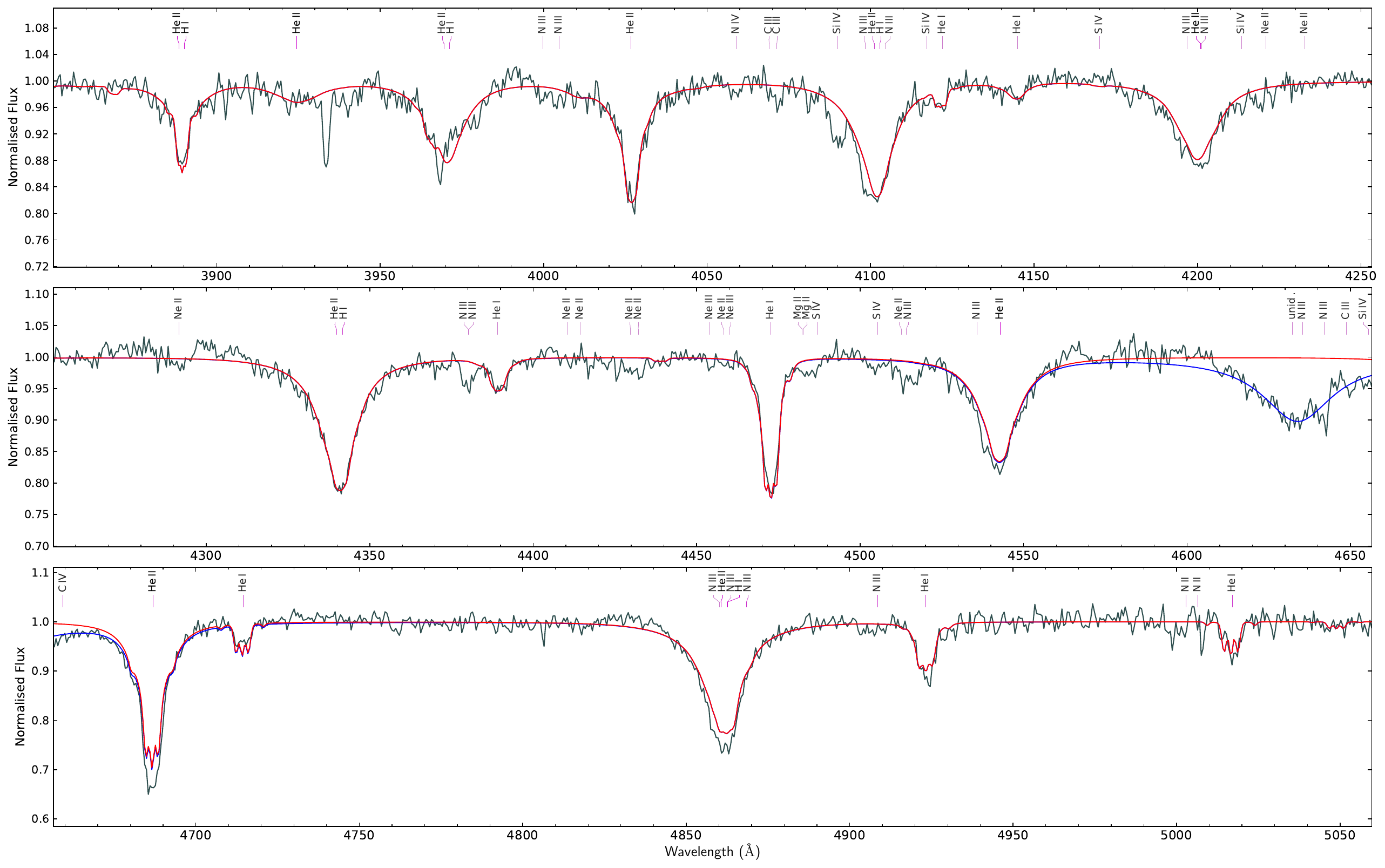}
        \caption{The co-added RSS spectrum of \J123\ (black) compared to the best two-component model (red). 
 }
        \label{fig:J123_full}
 \end{figure}

 \begin{figure*}
        \centering
        \includegraphics[width=24.2cm,angle=90]{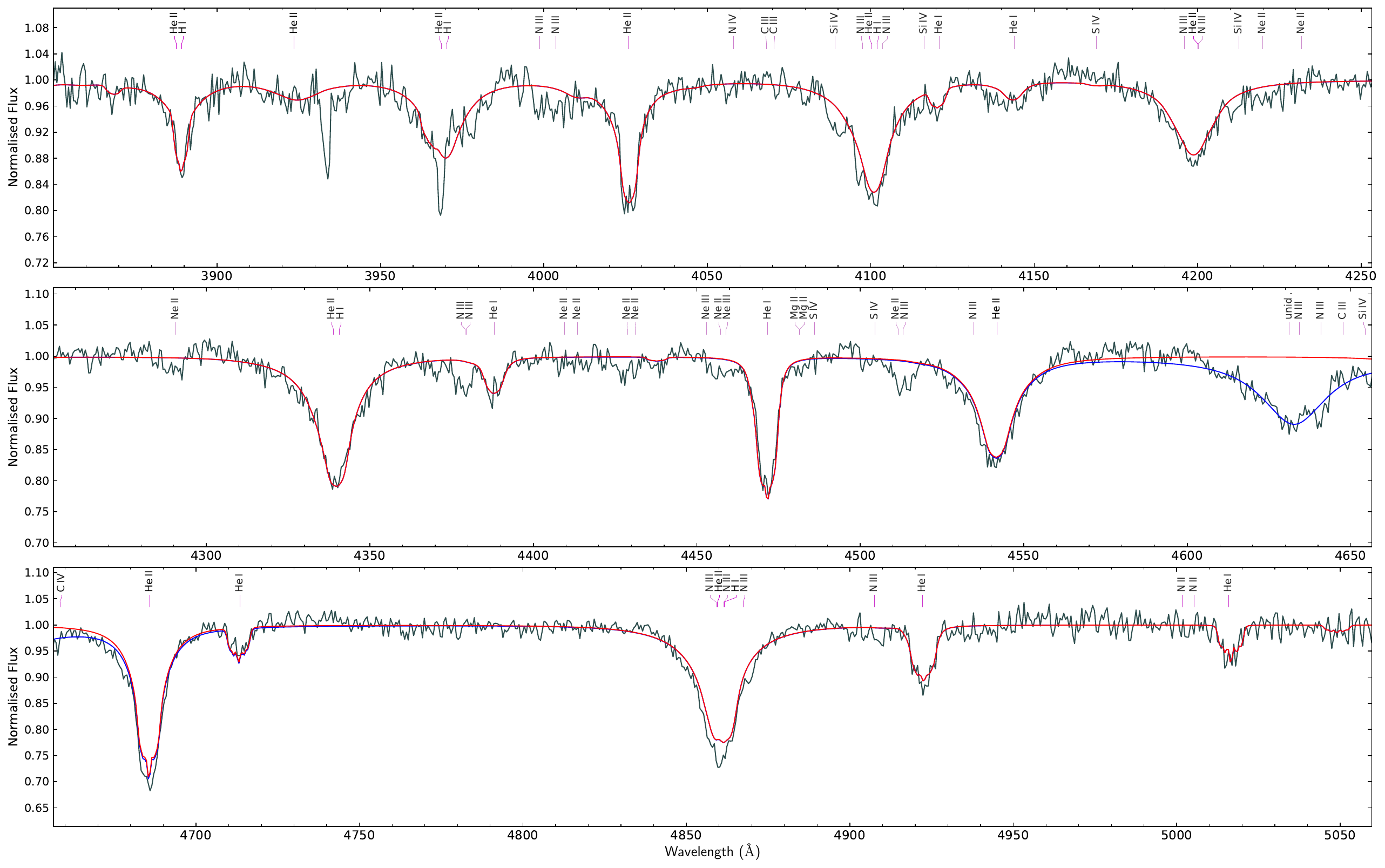}
 \caption{The co-added RSS spectrum of \J125\ (black) compared to the best two-component model (red).}
        \label{fig:J125_full}
 \end{figure*}

 \begin{figure*}
        \centering
        \includegraphics[width=24.2cm,angle=90]{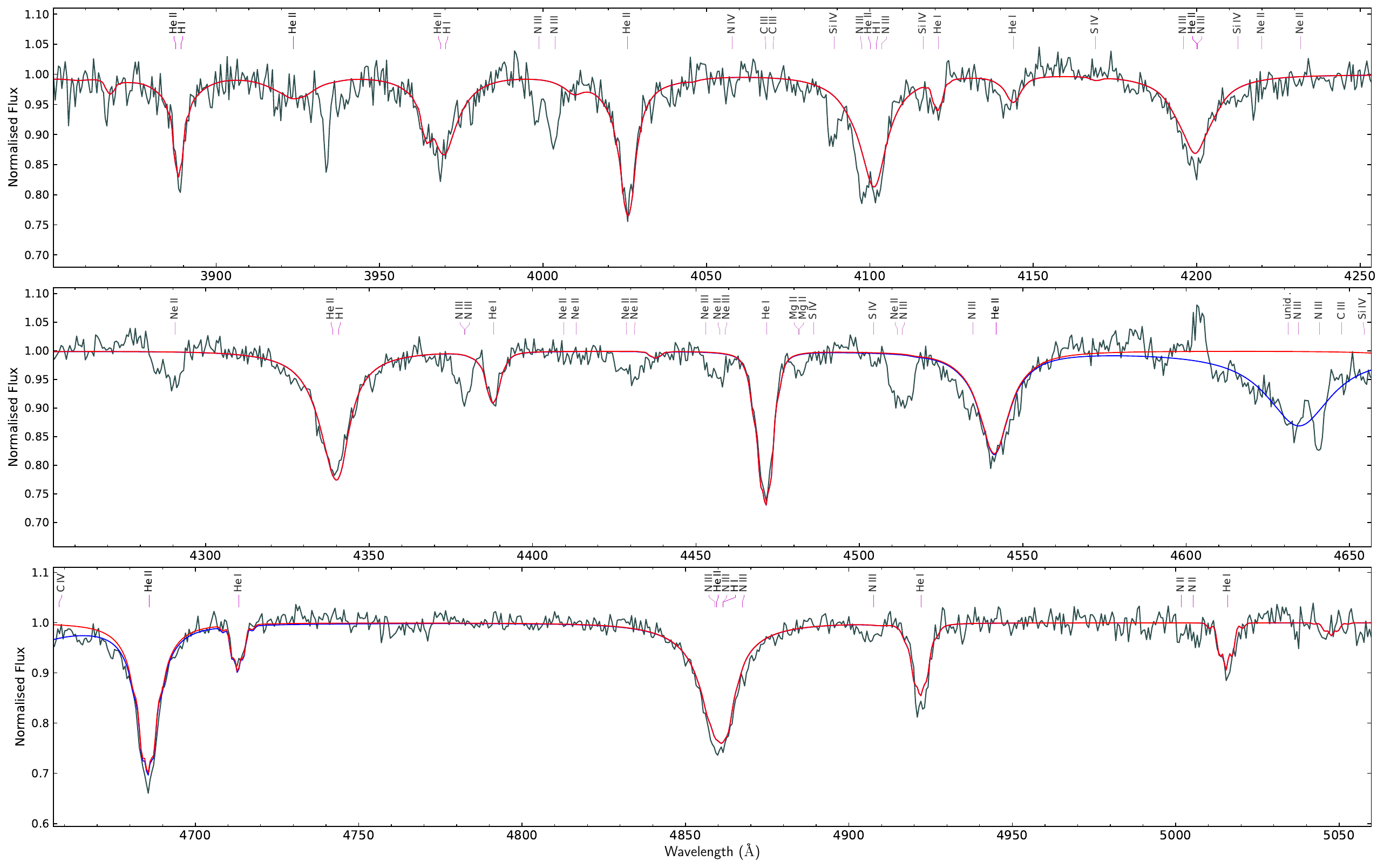}
        \caption{The co-added RSS spectrum of \J144\ (black) compared to the best two-component model (red). }
        \label{fig:J144_full}
 \end{figure*}

 \begin{figure*}
        \centering
        \includegraphics[width=0.8\textwidth]{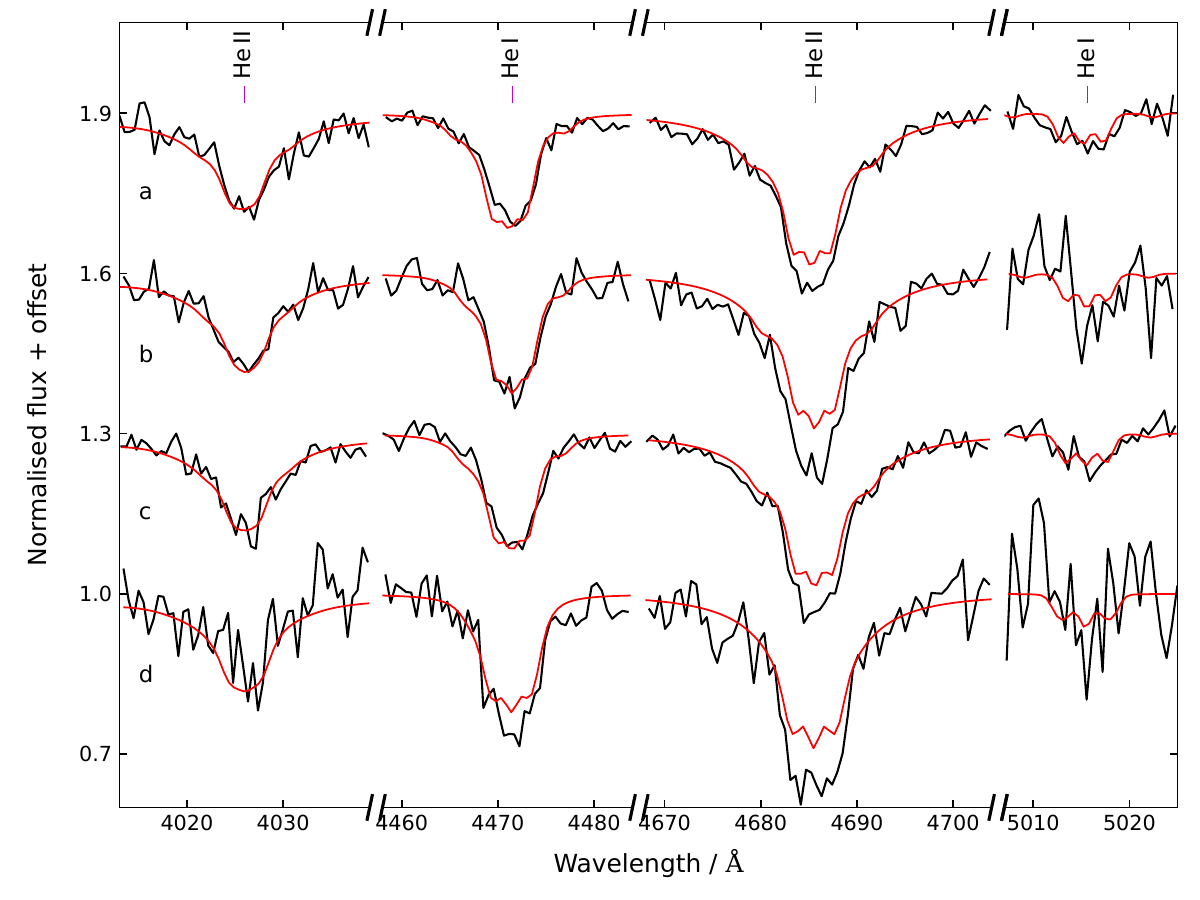}
        \caption{Best-fits (red) to individual spectra (black) of \J123. Each spectrum is identified by a letter as listed in Table \ref{tab:spec_mag}. Only the most important lines are shown. The models were allowed to vary in magnetic field strength, component contribution, and radial velocity. Other atmospheric parameters were kept the same for all spectra. }
        \label{fig:spec_var_1}
 \end{figure*}

  \begin{figure*}
        \centering
        \includegraphics[width=0.8\textwidth]{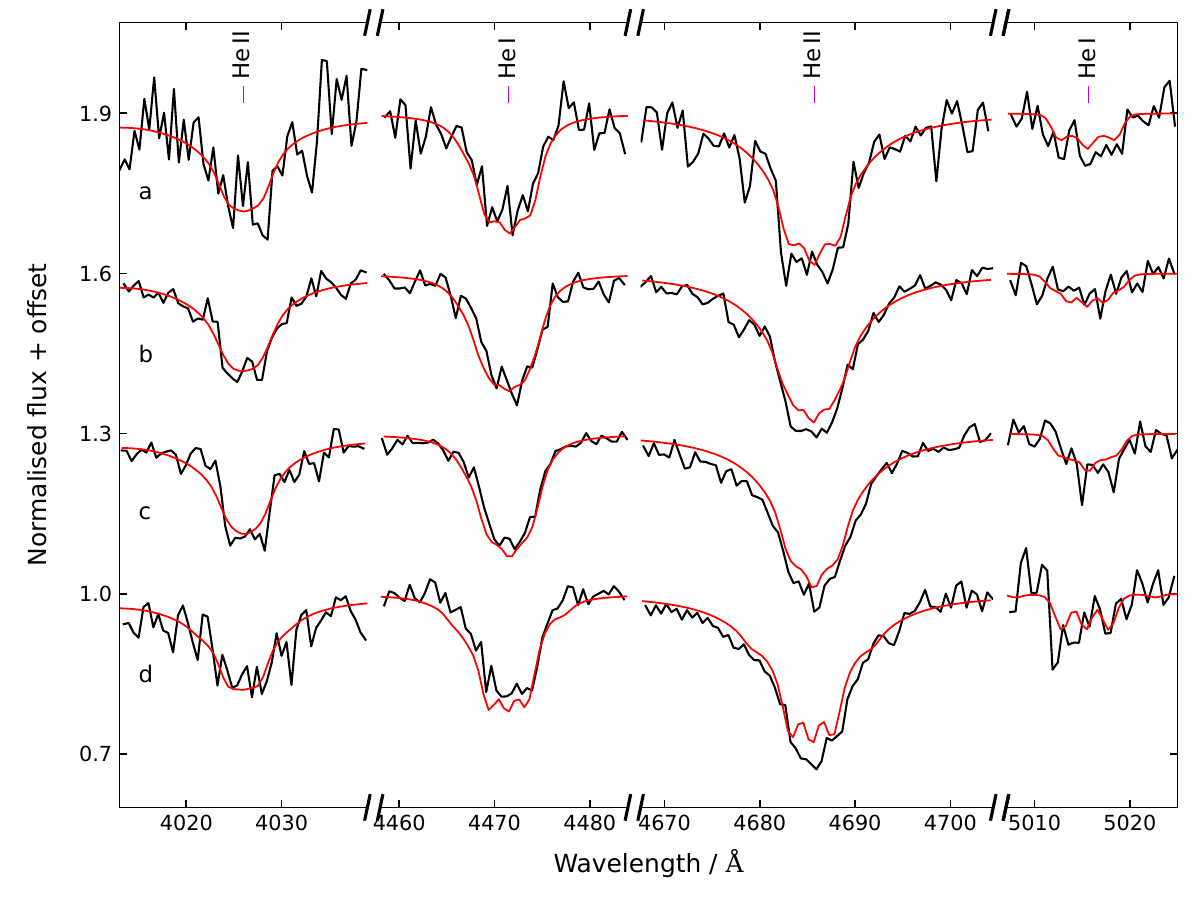}
        \caption{Like Fig.\ \ref{fig:spec_var_1}, but for \J125.}
        \label{fig:spec_var_2}
 \end{figure*}

 \begin{figure*}
        \centering
        \includegraphics[width=0.8\textwidth]{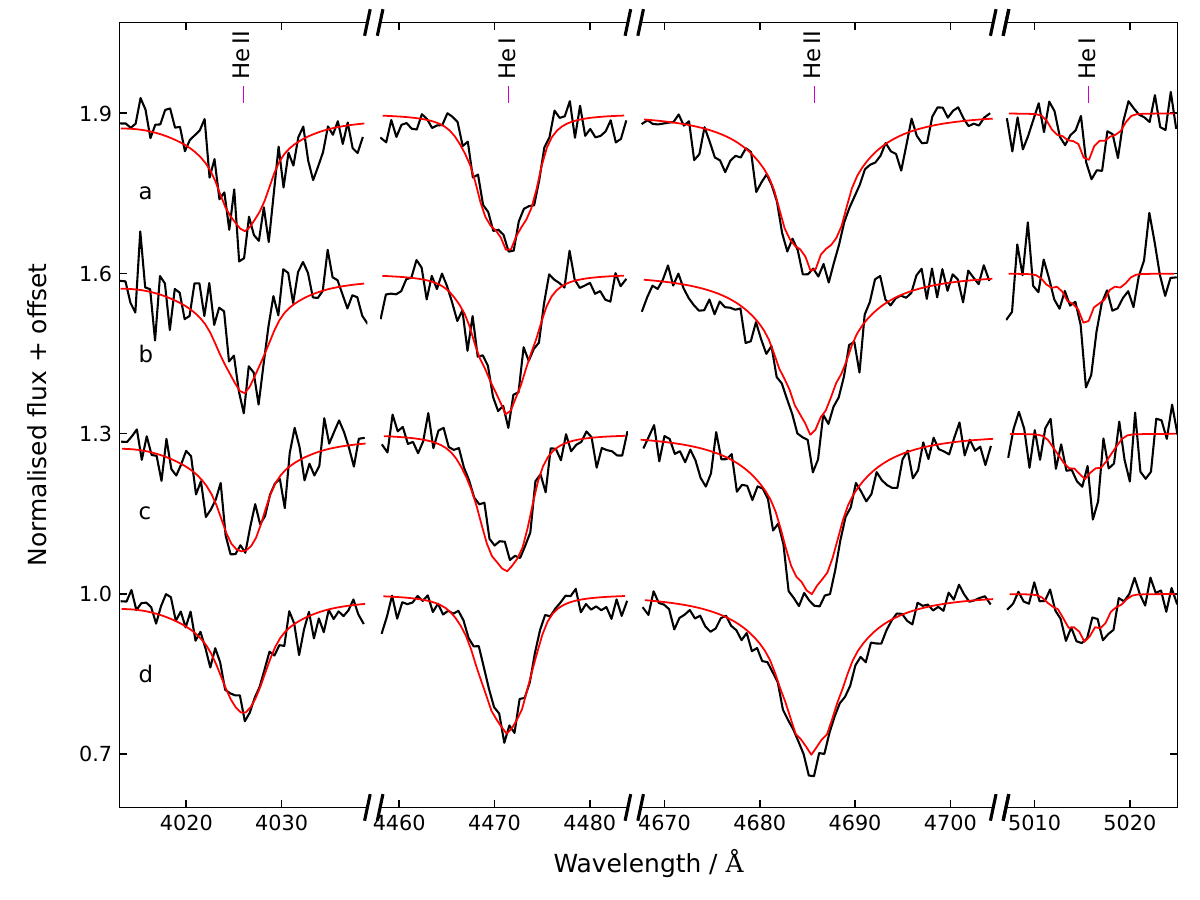}
        \caption{Like Fig.\ \ref{fig:spec_var_1}, but for \J144.}
        \label{fig:spec_var_3}
 \end{figure*}

\begin{figure*}
        \centering
        \includegraphics[width=0.8\textwidth]{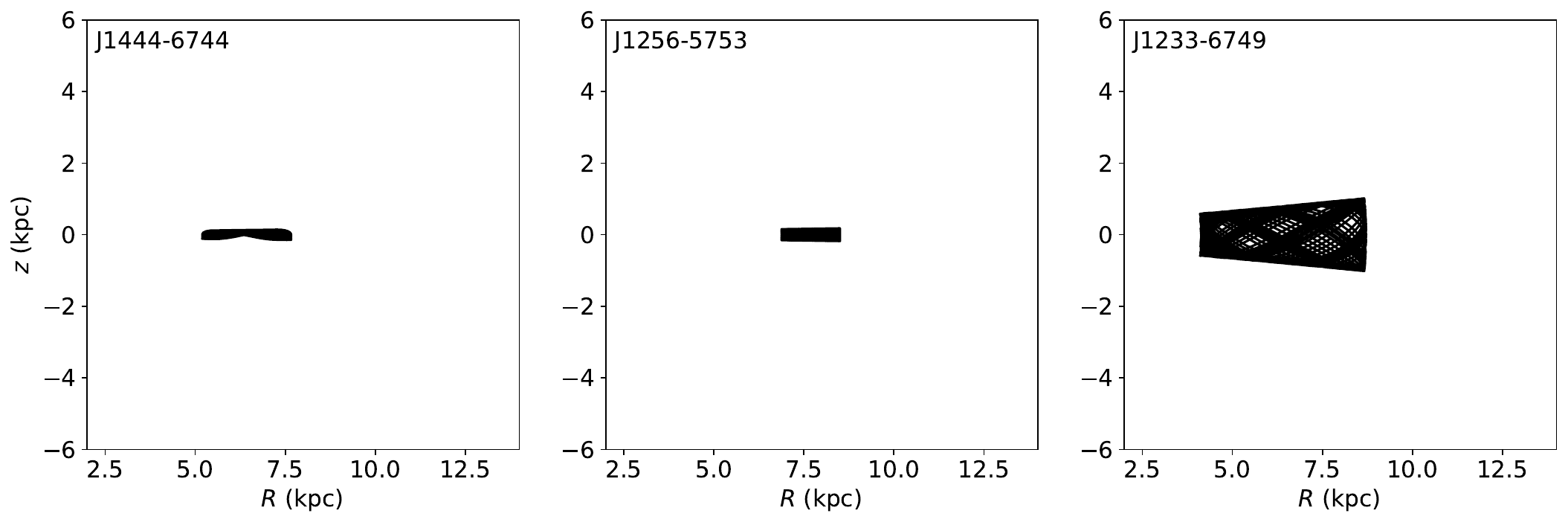}
        \includegraphics[width=0.99\textwidth]{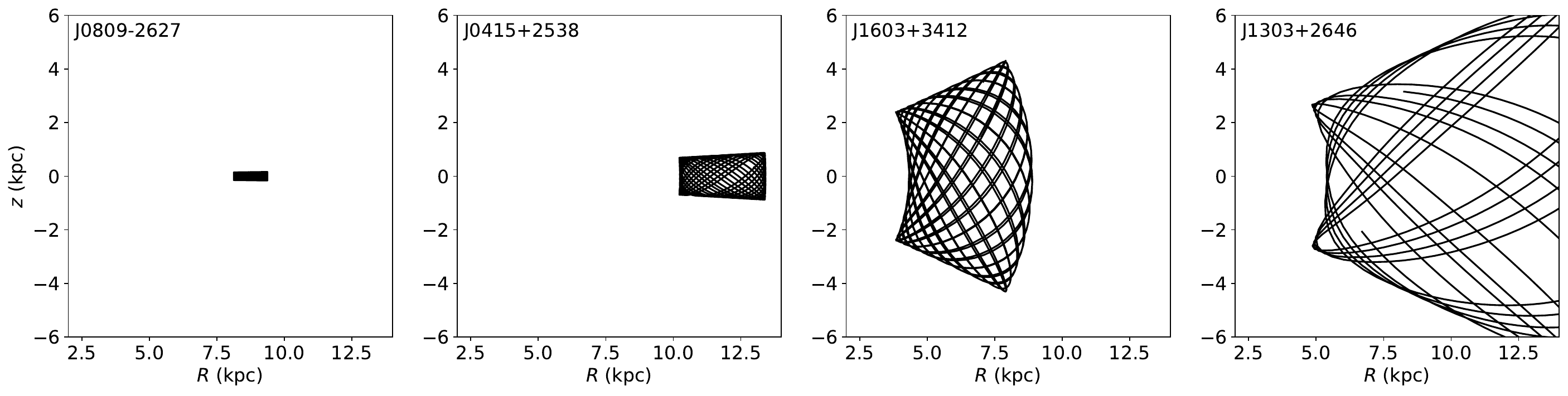}
        \caption{Our \texttt{galpy} orbits for the confirmed magnetic He-sdOs, showing motion in the $R$ versus $z$ plane integrated over 3 Gyrs from their current position. The top row shows the stars discussed in this paper while the bottom row shows the previously discovered magnetic He-sdO stars. 
    }
        \label{fig:galpy}
 \end{figure*}

  \begin{figure*}
        \centering
        \includegraphics[width=0.48\textwidth]{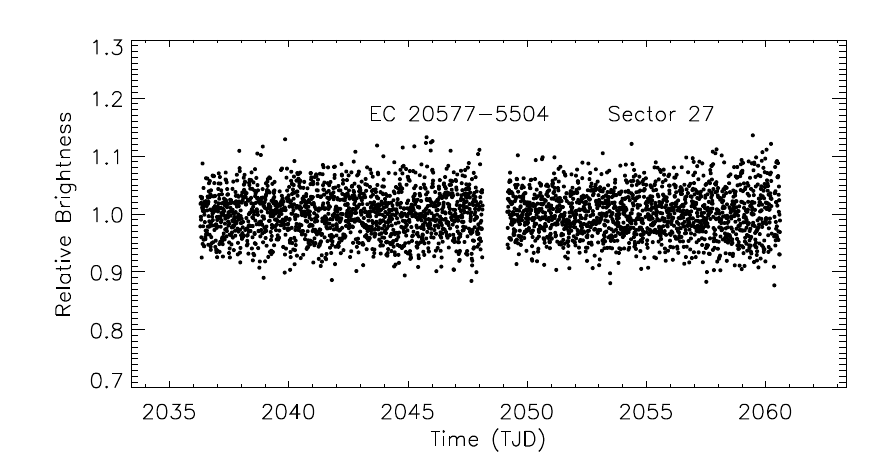}
        \includegraphics[width=0.48\textwidth]{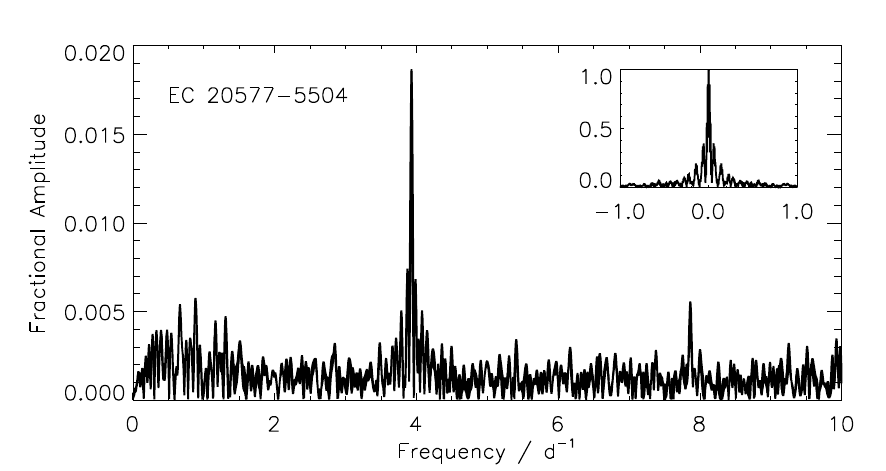}\\
        \includegraphics[width=0.48\textwidth]{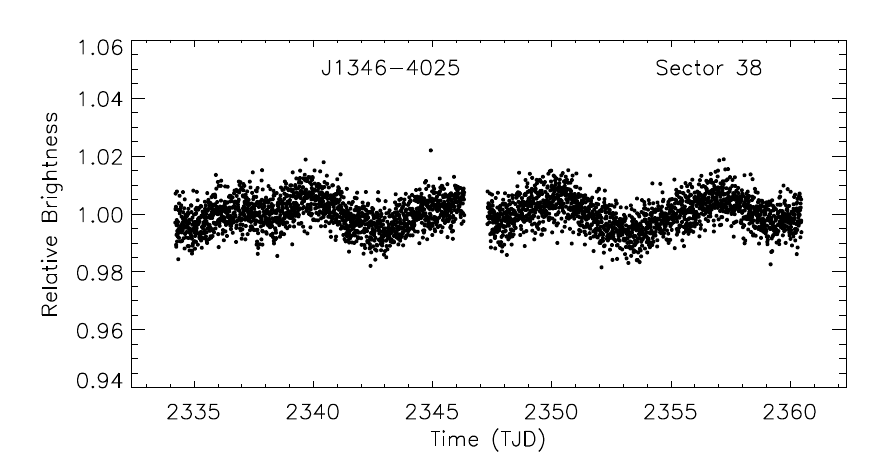}
        \includegraphics[width=0.48\textwidth]{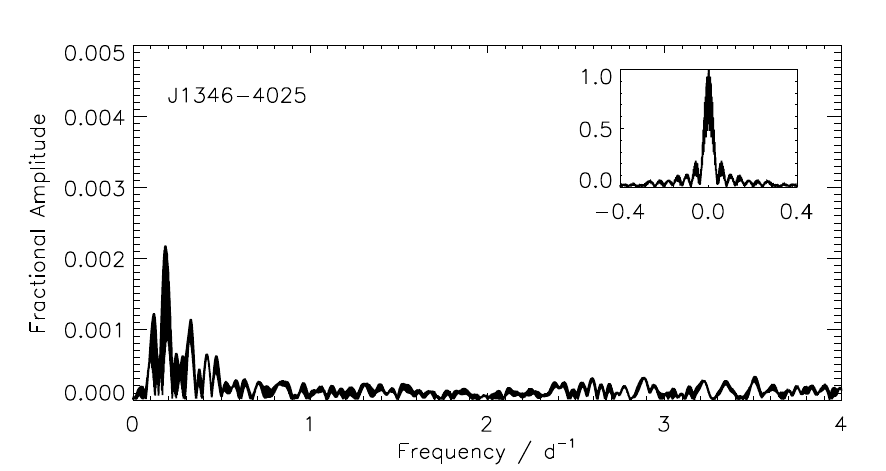}\\
    \includegraphics[width=0.48\textwidth]{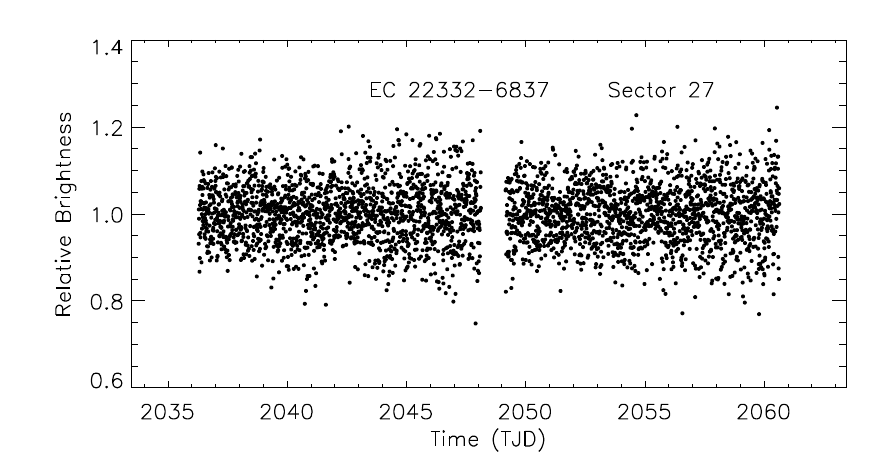}
        \includegraphics[width=0.48\textwidth]{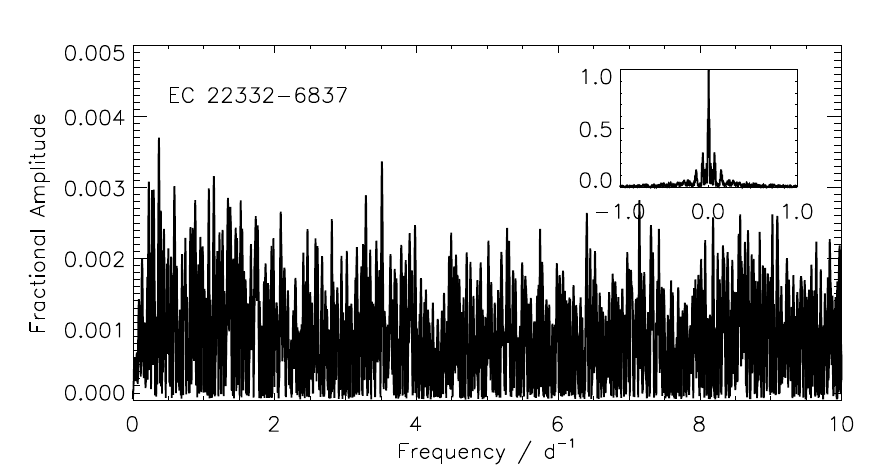}\\
        \caption{TESS light curves (left) and frequency-amplitude spectra (right) for three He-sdOs that show the 4630\,\AA\ feature. Light curves in units of relative brightness are shown for the best single sector in each case (as labelled). Times are in Julian days after start of the TESS mission. The amplitude spectra in units of relative brightness were computed from all available sectors.
    The inset shows the spectral window function at the same horizontal scale. 
    Note that the light curve of EC20577$-$5504 is likely contaminated by a close-by star. 
 }
        \label{fig:TESS}
 \end{figure*}

 \begin{table*}
\centering
\caption{Galactic velocities of the new and the known magnetic He-sdOs and orbital parameters. 
}
\label{tab:galpy}
\setlength{\tabcolsep}{4pt}
\begin{tabular}{l r@{$\pm$}l r@{$\pm$}l r@{$\pm$}l r@{$\pm$}l r@{$\pm$}l r@{$\pm$}l r@{$\pm$}l r@{$\pm$}l c}
\toprule
\toprule
Star & \multicolumn{2}{c}{$U$ / \kmsec} & \multicolumn{2}{c}{$V$ / \kmsec} & \multicolumn{2}{c}{$W$ / \kmsec} & 
    \multicolumn{2}{c}{$L_z$ / kpc\,\kmsec} &
    \multicolumn{2}{c}{$e$} & \multicolumn{2}{c}{$R_\mathrm{apo}$ / kpc} &
    \multicolumn{2}{c}{$R_\mathrm{peri}$ / kpc} & \multicolumn{2}{c}{$z_{\rm max}$ / kpc} & pop \\
\midrule
J1233-6749 & $-84.5$ & $3.4$ & $175.0$ & $8.6$ & $-51.0$ & $1.0$ & $1374$ & $65$ & $0.36$ & $0.03$ & $8.72$ & $0.01$ & $4.12$ & $0.30$ & $1.02$ & $0.04$ & TK\\[0.5mm]
J1256-5753 & $-52.1$ & $3.0$ & $242.0$ & $6.0$ & $-4.0$ & $0.6$ & $1858$ & $47$ & $0.11$ & $0.01$ & $8.51$ & $0.11$ & $6.90$ & $0.28$ & $0.18$ & $0.01$ & TH\\[0.5mm]
J1444-6744 & $-49.3$ & $3.3$ & $226.2$ & $4.5$ & $16.2$ & $0.7$ & $1536$ & $38$ & $0.19$ & $0.02$ & $7.66$ & $0.04$ & $5.20$ & $0.21$ & $0.15$ & $0.01$ & TH\\[0.5mm]
\midrule
J0415+2538 & $-18.7$ & $5.4$ & $256.6$ & $0.9$ & $16.6$ & $1.8$ & $2695$ & $40$ & $0.13$ & $0.01$ & $13.44$ & $0.30$ & $10.23$ & $0.17$ & $0.88$ & $0.07$ & TH\\[0.5mm]
J0809$-$2627 & $-29.5$ & $0.8$ & $231.2$ & $1.0$ & $11.2$ & $0.1$ & $2099$ & $10$ & $0.07$ & $0.01$ & $9.37$ & $0.05$ & $8.13$ & $0.05$ & $0.16$ & $0.00$ & TH\\[0.5mm]
J1303+2646 & $-110.1$ & $5.5$ & $363.0$ & $6.2$ & $49.7$ & $6.1$ & $1729$ & $50$ & $0.53$ & $0.05$ & $17.50$ & $1.93$ & $5.39$ & $0.18$ & $7.79$ & $1.83$ & H\\[0.5mm]
J1603+3412 & $-25.7$ & $1.2$ & $255.4$ & $2.6$ & $101.0$ & $7.3$ & $1173$ & $130$ & $0.35$ & $0.07$ & $9.01$ & $0.32$ & $4.36$ & $0.51$ & $4.31$ & $0.68$ & TK\\[0.5mm]
\bottomrule
\end{tabular}
\tablefoot{
The table lists the $z$-component of angular momentum ($L_z$), eccentricity (e), galactic apocentre ($R_{\rm apo}$), galactic pericentre ($R_{\rm peri}$) and maximum vertical amplitude ($z_{\rm max}$). The left-handed system for the Galactic velocity components is used here, where $U$ is positive towards the Galactic Centre, $V$ is in the direction of Galactic rotation and $W$ is positive towards the North Galactic Pole. The last column shows the population classification as described in \citet{PhilipMonai2024}, where TH $\equiv$ thin disk, TK $\equiv$ thick disk, and H $\equiv$ halo. 
}

\end{table*}

\end{onecolumn}

\end{appendix}

\end{document}